\documentclass{article}
\usepackage[utf8]{inputenc}
\usepackage{graphicx}
\usepackage{multirow}
\usepackage{placeins}
\usepackage{capt-of}
\usepackage{booktabs}
\usepackage{varwidth}
\usepackage{xcolor}
\usepackage{float}
\usepackage[T1]{fontenc}
\usepackage{authblk}
\usepackage{subfig}

\title{Low field electron mobility in $\alpha-Ga_{2}O_{3}$: An ab-initio approach}
\author[1]{Ankit Sharma \thanks{ankitsha@buffalo.edu}}
\author[1]{Uttam Singisetti}
\affil[1]{Department of Electrical Engineering, University at Buffalo}

\begin{document}

\maketitle

\section{Abstract}
The $\alpha$ phase of $Ga_{2}O_{3}$ is an ultra-wideband semiconductor with potential power electronics applications. In this work, we calculate the low field electron mobility in $\alpha-Ga_{2}O_{3}$ from first principles. The 10 atom unit cell contributes to 30 phonon modes and the effect of each mode is taken into account for the transport calculation.  The phonon dispersion and the Raman spectrum are calculated under the density functional perturbation theory formalism and compared with experiments. The IR strength is calculated from the dipole moment at the $\Gamma$ point of the Brillouin zone. The electron-phonon interaction elements (EPI) on a dense reciprocal space grid is obtained using the Wannier interpolation technique. The polar nature of the material is accounted for by interpolating the non-polar and polar EPI elements independently as the localized nature of the Wannier functions are not suitable for interpolating the long-range polar interaction elements. For polar interaction the full phonon dispersion is taken into account. The electron mobility is then calculated including the polar, non-polar and ionized impurity scattering. 

\section{Introduction}
Gallium oxide ($Ga_{2}O_{3}$) has emerged as a promising candidate for power electronics, RF and UV optoelectronic applications due to its large bandgap ($4.3-5.3 eV$). It is known to exist in 6 polymorphs of $\alpha$, $\beta$, $\gamma$, $\delta$, $\epsilon$ \cite{ref1} and $\kappa$  \cite{ref2}. Of the known polymorphs, $\beta$ phase is the most thermodynamically stable phase \cite{ref3} and hence it is also the most extensively studied phase. It has a bandgap of about 4.7-4.9 eV \cite{ref4} which results in a very high critical field strength (~8 MV/cm) and a high BFoM \cite{ref4,ref5} compared to SiC and GaN which currently dominate the high power device market. Lateral transistor with breakdown voltage of 1.8 kV \cite{ref6}, Schottky barrier diodes \cite{ref7,ref8} and vertical transistors \cite{ref9} have been demonstrated using the $\beta-Ga_{2}O_{3}$. In addition, (AlGaO/GaO) heterostructures \cite{ref10} are also explored to achieve high electron mobility, since the $\beta$ phase is known to have low bulk electron mobilities owing to high polar optical phonon scattering \cite{ref11}.
On the contrary, $\alpha-Ga_{2}O_{3}$ has received comparatively less attention as it is thermodynamically semi-stable making its synthesis challenging which requires high growth temperatures $(>430^{0}C)$ and pressures \cite{ref12}. However, recently thin films were successfully grown under low temperature conditions on sapphire substrate using atomic layer deposition technique \cite{ref12} paving way for the low cost substrates. High quality $\alpha-Ga_{2}O_{3}$ films are also grown by lateral epitaxial overgrowth techniques by HVPE with low background densities \cite{ref35}. It is structurally analogous to corundum $\alpha-Al_{2}O_{3}$ ($R3-c$) and thus can be potentially alloyed with corundum to fabricate tunable bandgap structures and with materials such as $Cr_{2}O_{3}$ and $Fe_{2}O_{3}$ for magnetoelectric applications. It has a higher bandgap (~5.3 eV) \cite{ref13} when compared to the $\beta$ phase, hence a better performance is expected for high power device applications.
\newline
\newline
There have been extensive studies on the mechanism of electron transport in $\beta-Ga_{2}O_{3}$ under both the low and high field conditions \cite{ref11,ref14}. In contrast, very few reports exist on understanding the fundamental transport mechanism in $\alpha$-phase. With the maturity in the growth of the $\alpha$ phase, it is important to investigate the transport phenomena in $\alpha-Ga_{2}O_{3}$ to design better devices and optimize performance. There are few reports on the electronic structure and the optical properties of the $\alpha$ phase \cite{ref3} and none on the theoretical study of electron transport.
\newline
\newline
$\alpha-Ga_{2}O_{3}$ is a polar semiconductor due to the electronegativity difference of the constituent atoms and hence there arises a need to investigate two types of electron-phonon scattering i.e. the polar optical phonon scattering which is the most dominant scattering mechanism in$\beta-Ga_{2}O_{3}$ and the non-polar deformation potential scattering. A large unit cell results in multiple phonon modes contributing to the scattering process making the analysis challenging. In this work, we calculate from the first principles the dominant scattering mechanism among the long range polar scattering, the short range non polar scattering and the ionized impurity scattering. A mode-by-mode analysis of the dominant scattering mechanism limiting the low field electron mobility is also presented.

\section{Theory and Methods}
\subsection{Computational Methods}
\begin{figure}
\includegraphics[width=1.2\linewidth]{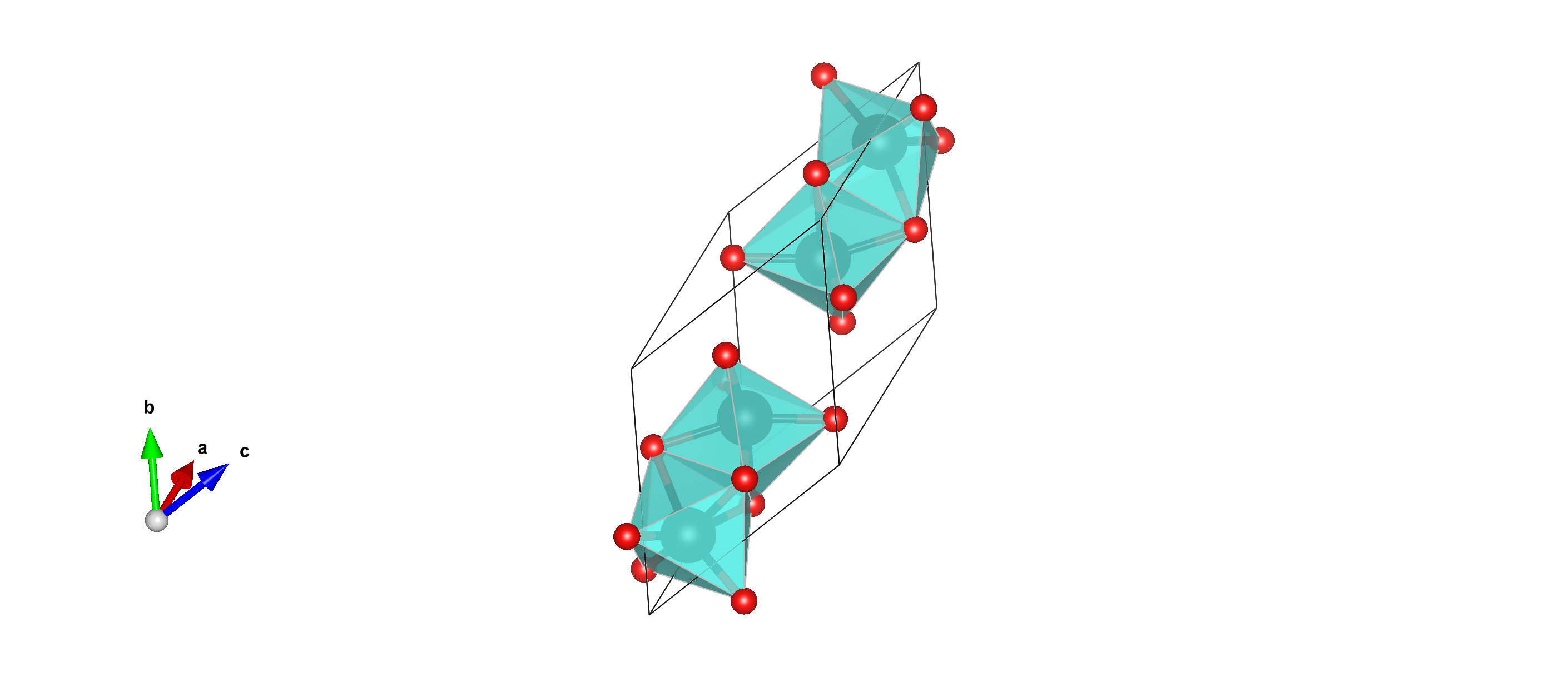}
\caption{The 10 atom conventional cell of $\alpha-Ga_{2}O_{3}$ where unlike $\beta-Ga_{2}O_{3}$, all the Gallium atoms have the same co-ordination}
\label{fig:PrimCell}
\end{figure}

We use Quantum Espresso \cite{ref17,ref18}, a planewave pseudopotential based DFT package for our calculation. All the calculations in the present work is done under the Local Density Approximation (LDA) for the exchange-correlation functional. In our calculations, we have used a 10 atom rhombohedral primitive cell which has 4 equivalent Ga (Gallium) atoms in distorted octahedral sites and 6 O (Oxygen) atoms as shown in the $fig\ref{fig:PrimCell}$ along with the first Brillouin zone generated using XCrysden \cite{ref15} and Vesta  \cite{ref16} whereas the conventional cell has 30 atoms.

We first start with the structural relaxation after accounting for the convergence in the planewave energy cutoff and the reciprocal space grid. We have used the cutoff energy for planewave expansion of $1088  eV$, charge density cutoff of $4352 eV$ and a $\Gamma$ centered grid of size $6\times6\times6$ by checking for the convergence of the total energy. We have taken the Gallium 3d orbitals as the part of the valence bands when choosing the pseudopotential. After the structural relaxation, we perform the atomic relaxation to minimize the forces on the atoms ($ < 1e^{-4 }Ry/Bohr$) which otherwise might give spurious results due to the structural instability during the lattice dynamics calculations. The relaxed structural parameters given in $table\ref{table:LatPara}$ \cite{ref34} are dependent on the type of the pseudopotential used during the self-consistent run.

\FloatBarrier
\begin{table}[h]
\centering
\begin{tabular}{|c|c|c|}
\hline
\multicolumn{3}{|c|}{$\textbf{Lattice Parameters}$}                                           \\ \hline
\multicolumn{3}{|l|}{}                                                             \\ \hline
$Property$                                                  & $This Work$ & $Experiment^{[20]}$ \\ \hline
a ($A$)                                           & 4.9876    & 4.983      \\ \hline
\begin{tabular}[c]{@{}c@{}}c ($A$)\\ \end{tabular} & 13.3593   & 13.433     \\ \hline
\multicolumn{3}{|l|}{}                                                             \\ \hline
\multicolumn{3}{|c|}{Fractional Coordinates}                                       \\ \hline
z (Ga)                                                    & 0.3565    & 0.3554     \\ \hline
x (O)                                                     & 0.3018    & 0.3049     \\ \hline
\end{tabular}
\caption{The lattice parameters obtained after the structural and atomic relaxation, showing close match with the experimentally observed values}
\label{table:LatPara}
\end{table}
\FloatBarrier

\subsection{Electronic Structure}
\begin{figure}[H]
\centering
\includegraphics[width=\linewidth]{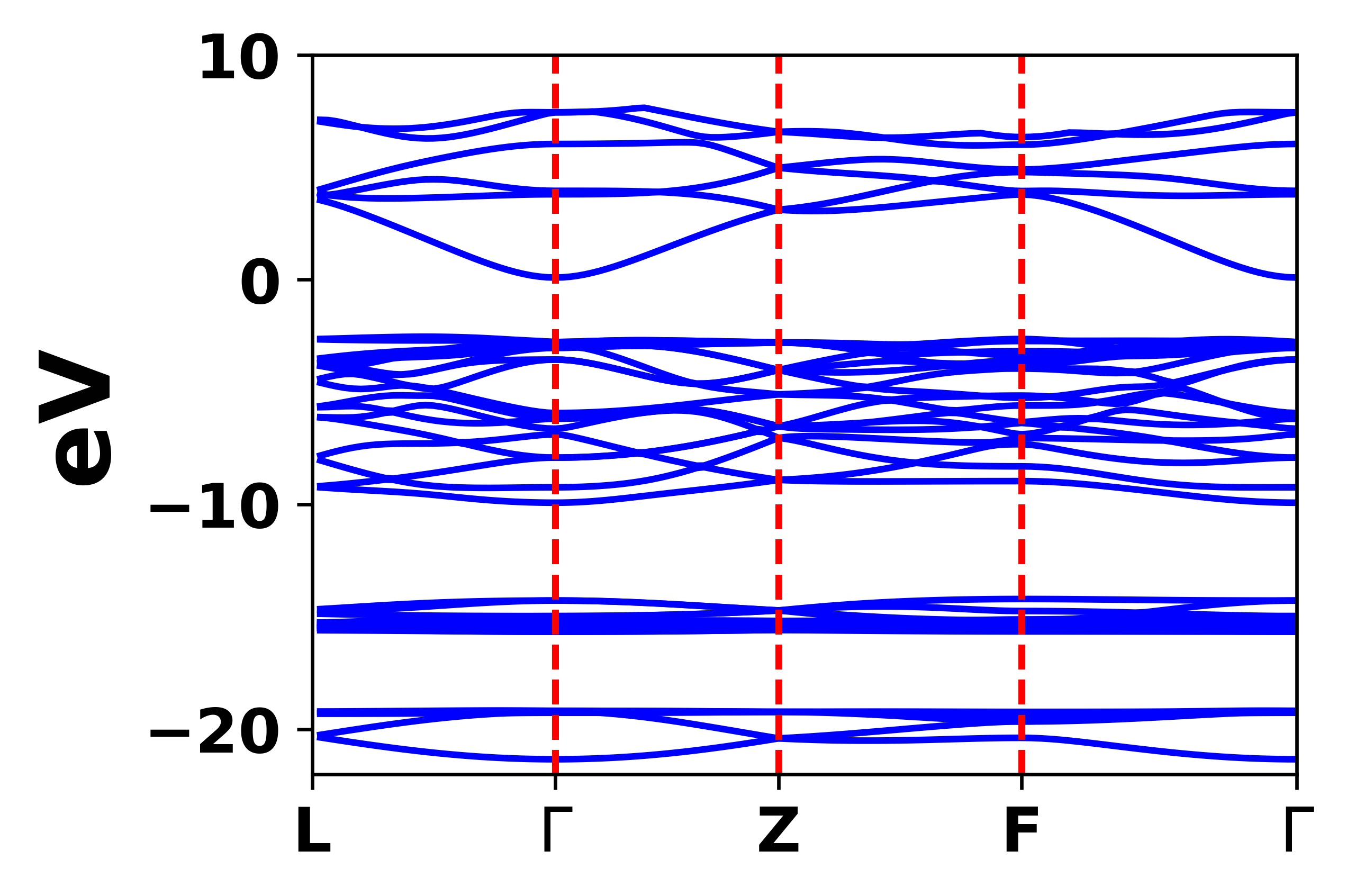}
\caption{The LDA electronic structure using the DFT calculation in the IBZ. Primarily an indirect bandgap with valence band maxima at point F(0.5 0.5 0) and conduction band minima at $\Gamma$ point, but the difference between the two band gaps in very less}
\label{fig:Band}
\end{figure}

\begin{figure}[H]
  \centering
  \subfloat[Dynamical matrix]{\includegraphics[width=0.5\textwidth]{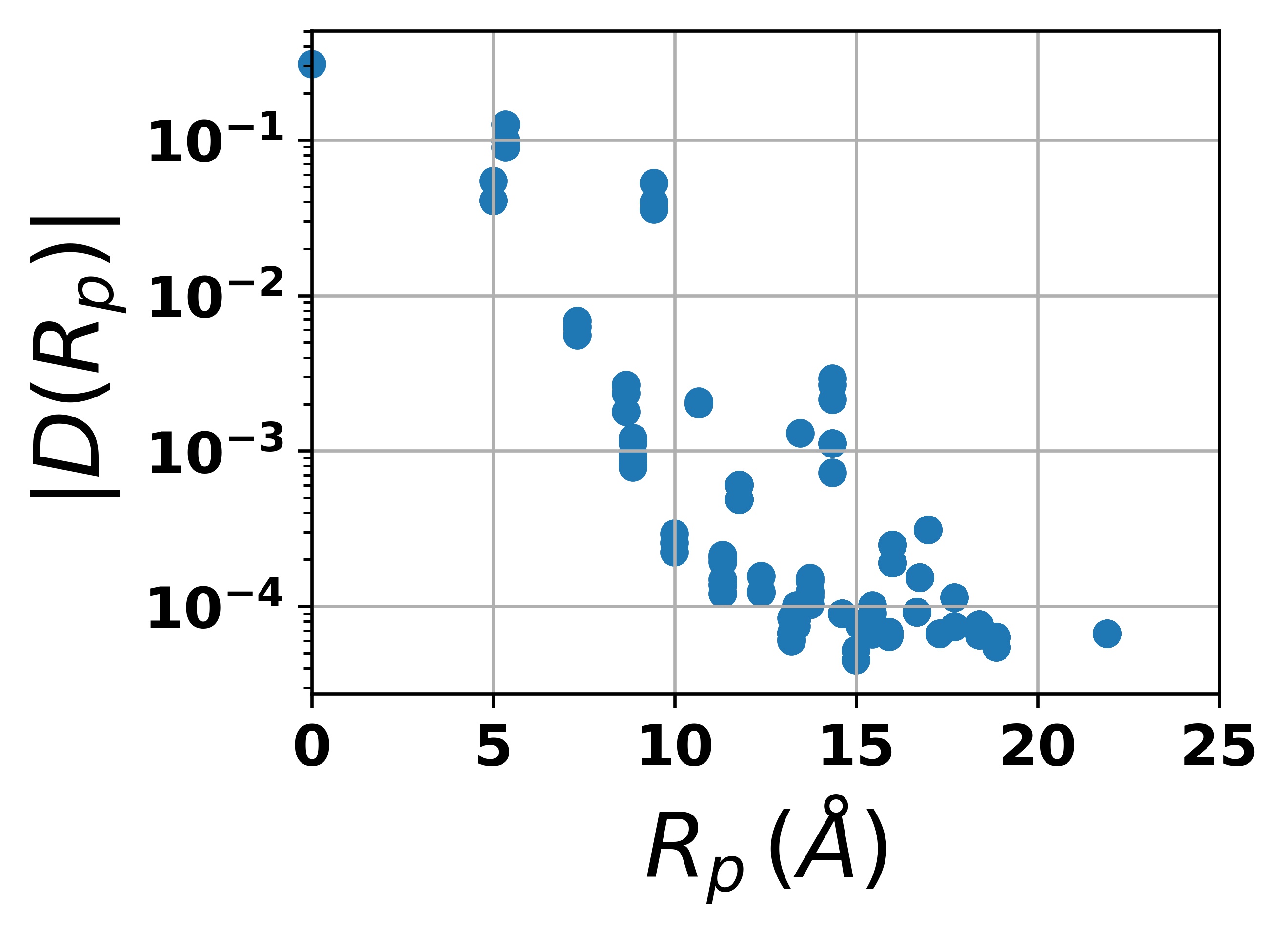}\label{fig:f1}}
  \hfill
  \subfloat[Hamiltonian]{\includegraphics[width=0.5\textwidth]{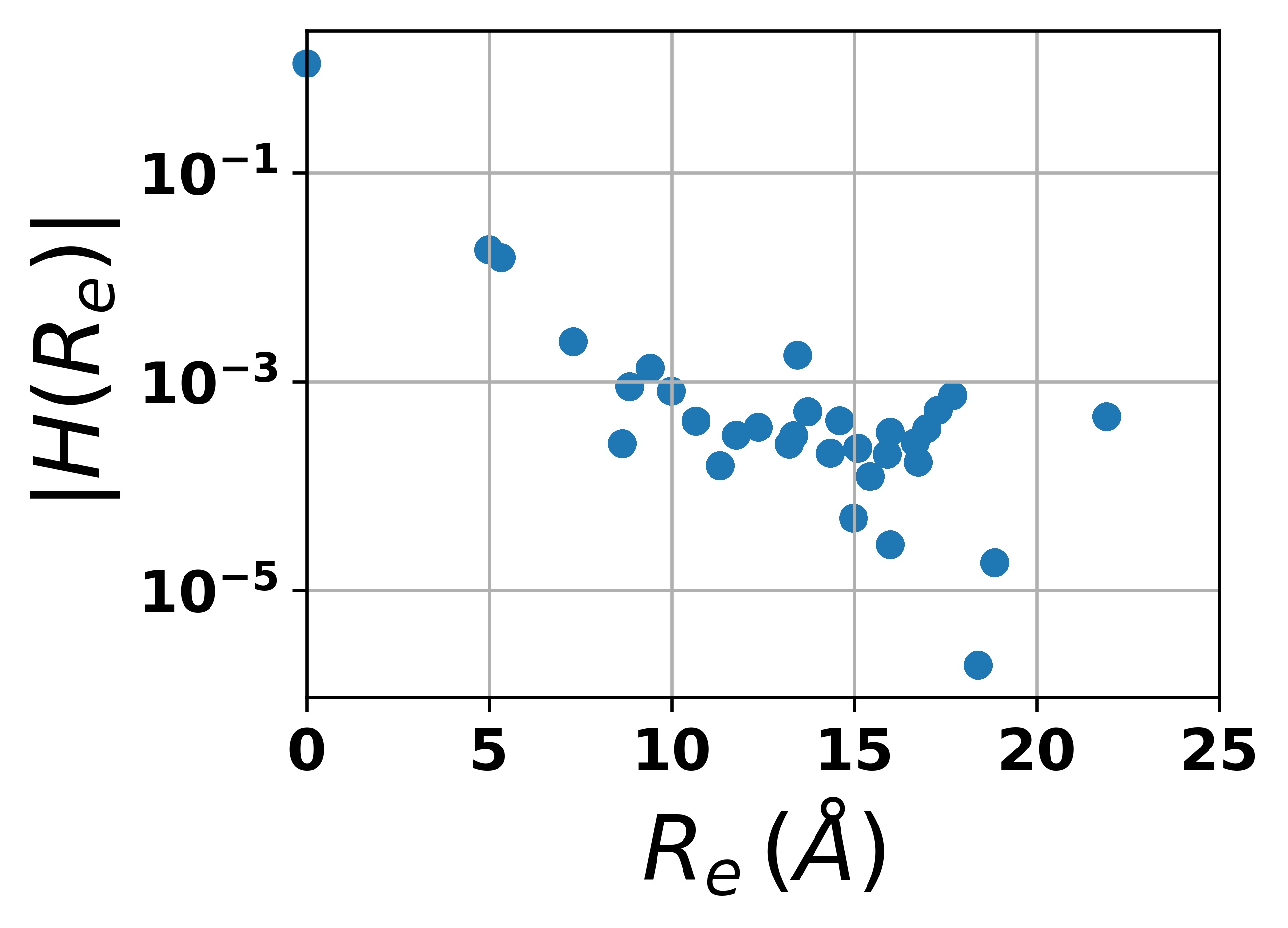}\label{fig:f2}}
  \caption{The decay of the dynamical matrix in real space representing the quality of localization using the Wannier Functions. b). The decay of the Hamiltonian in the real space}
\end{figure}

The electronic structure is calculated under the local density approximation, as such it underestimated the band gap which is a well known problem arising mainly due to the error in the self-interaction correction by the exchange-correlation functional. The bandgap problem can be overcome by use of higher order XC (exchange-correlation) functionals, and there are some results of this in the literature which estimates the $\alpha-Ga_{2}O_{3}$ indirect bandgap of $5.03 eV$ and a direct bandgap of $5.08 eV$ \cite{ref3}. Since the low field electron transport is not impacted by the bandgap, we use LDA band structure in our calculations as shown in the $fig\ref{fig:Band}$. After computing the converged charge density on a coarse reciprocal space grid of size $6\times6\times6$, we employ the technique of Wannier interpolation \cite{ref19,ref20,ref21} to obtain the electronic structure on a fine grid. The idea behind the Wannier interpolation is to do a unitary rotation of the Hamiltonian from the Bloch space to the localized Wannier space and then rotating it back to the fine grid in the Bloch space. This is done using the maximally localized Wannier functions where the Hamiltonian in the Wannier space is given as:
\begin{equation}
H_{R_{e},R_{e'}}^{el} = \sum_{\textbf{k}}e^{-i\textbf{k}.(R_{e}-R_{e'})}U_{\textbf{k}}^{\dagger}H_{\textbf{k}}^{el}U_{\textbf{k}}
\end{equation} 

where $H^{el}_{k}$ is the electronic Hamiltonian in the coarse Bloch space, $R_{e}$ is the Wigner-Seitz cell centers and $U$ are the unitary rotation matrices for band mixing and gauge transformation. During the interpolation using Wannier functions, we also check for the spatial localization of the electronic Hamiltonian and the phonon Dynamical matrix, the plots of which are shown in $fig\ref{fig:f1}$ and $fig\ref{fig:f2}$ respectively. The figures clearly shown qualitatively the spatial decay of the Hamiltonian and the Dynamical matrix in the real space. The plot of the partial density of states shows that the valence band is formed by the O 2p orbitals and the conduction band by the Ga 4s orbitals, similar to that observed in $\beta-Ga_{2}O_{3}$. The calculated effective mass is $\textbf{0.25*m0}$, which is comparable to the previous reported value \cite{ref3}. The effective mass is slightly lower than the $\beta$ phase (0.3) \cite{ref11}, which should help in increasing the electron mobility when compared to the $\beta$ phase.

\newpage
\subsection{Lattice Dynamics}
\begin{figure}
\centering
\includegraphics[width=\linewidth]{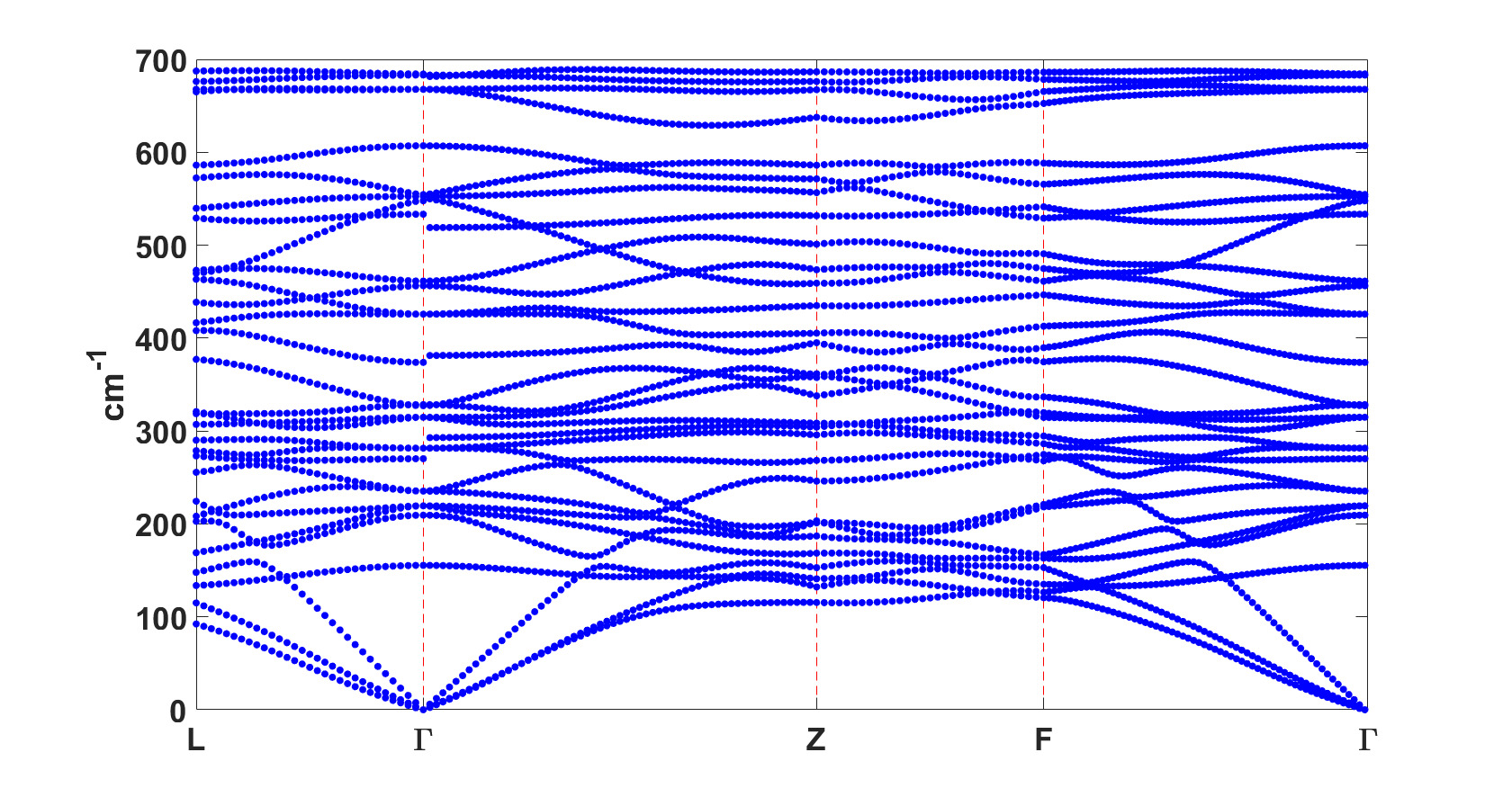}
\caption{The phonon dispersion obtained by the Fourier interpolation of the dynamical matrix with the inclusion of the non-analytical term for the macroscopic electric field resulting in the LO-TO split represented by the discontinuities at the $\Gamma$ point toward the Z and the L directions}
\label{fig:Phonon}
\end{figure}

We then use the density functional perturbation theory \cite{ref22}, a linear response theory as implemented in the Quantum Espresso package to calculate the lattice dynamical properties. A $6\times6\times6$ reciprocal space phonon wavevector grid is used for this calculation to obtain the polarization vectors and the perturbation in the potential due to the atomic vibration. To calculate the phonon dispersion along the directions of the electronic structure in the irreducible wedge of the Brillouin zone, the dynamical matrix is interpolated using the Fourier interpolation technique and then diagonalized on a dense grid. The resulting phonon spectrum is shown in the $fig\ref{fig:Phonon}$. The long range correction to the dynamical matrix arising from the coupling of the macroscopic field with the LO modes, which shifts the energy of the LO mode, is taken into account by explicitly adding the non-analytical term to the dynamical matrix before diagonalization. We can see the discontinuity (circled in red) in the phonon spectrum in the $fig\ref{fig:Phonon}$ at the $\Gamma$ point arising due to the split between the LO and TO modes. 

The polar nature of the material results in charges on the ions which maybe more than the oxidation state of the isolated ion. It is known as Born effective charge which is a tensor quantity. It is calculated at the $\Gamma$ point along with the high frequency dielectric tensor. Both these quantities are related to the atomic polarization. The DC dielectric constant is obtained using the LST (Lyddane-Sachs-Teller) relation. As shown in $table\ref{table:dielectric}$, the high frequency dielectric tensor is nearly isotropic, with the space group symmetry resulting in $\epsilon_{xx}=\epsilon_{yy}$ and a small anisotropy in $\epsilon_{zz}$ but the DC dielectric constant captures anisotropy arising due to the IR (infrared) mode polarization directions and the resulting direction dependent LO-TO split.

\FloatBarrier
\begin{table}[h]
\centering
\begin{tabular}{|c|c|c|c|c|c|}
\hline
\multicolumn{2}{|c|}{$\textbf{X}$} & \multicolumn{2}{c|}{$\textbf{Y}$} & \multicolumn{2}{c|}{$\textbf{Z}$} \\ \hline
$TO(meV)$ & $LO(meV)$ & $TO(meV)$ & $LO(meV)$ & $TO(meV)$ & $LO(meV)$ \\ \hline
27.2686 & 27.2748 & 27.2686 & 27.2748 & \multirow{2}{*}{33.4148} & \multirow{2}{*}{53.9783} \\ \cline{1-4}
40.5946 & 46.3179 & 40.5946 & 46.3179 & & \\ \hline
57.0628 & 67.7224 & 57.0628 & 67.7244 & \multirow{2}{*}{66.0624} & \multirow{2}{*}{83.6049} \\ \cline{1-4}
68.35 & 84.5809 & 68.35 & 84.5809 & & \\ \hline
\multicolumn{2}{|c|}{$\epsilon_{xx}^{\infty}$ = 4.62} & \multicolumn{2}{c|}{$\epsilon_{xx}^{\infty}$ = 4.62} & \multicolumn{2}{c|}{$\epsilon_{xx}^{\infty}$ = 4.465} \\ 
\multicolumn{2}{|c|}{$\epsilon^{DC}$ = 12.9816} & \multicolumn{2}{c|}{$\epsilon^{DC}$ = 12.9816} & \multicolumn{2}{c|}{$\epsilon^{DC}$ = 18.6645} \\ \hline
\end{tabular}
\caption{Anisotropic LO-TO split resulting in the large difference of the DC dielectric constant obtained using the LST relation. This is the consequence of the anisotropic polarization of the phonon modes where the x and y ($E_{u}$) direction shows equal split because of degeneracy with the maximum split in the z ($A_{2u}$) direction}
\label{table:dielectric}
\end{table}
\FloatBarrier

\begin{figure}[H]
\centering
\includegraphics[width=0.8\linewidth]{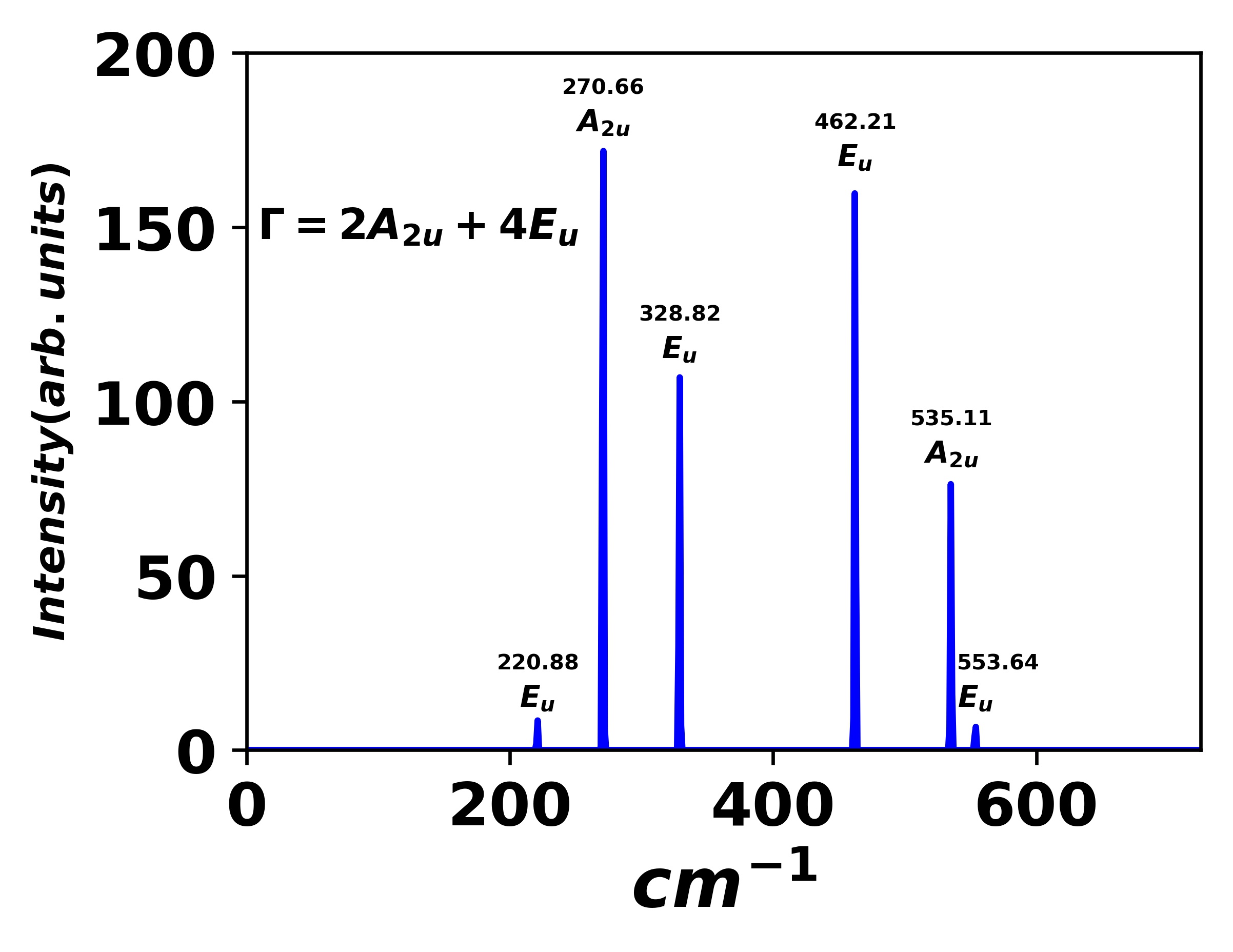}
\caption{The IR strength calculated using the product of the Born effective charge and the displacement pattern for the active modes. The $10^{th}$ mode is the strongest mode and is expected to contribute most to the polar optical scattering rate}
\label{fig:IR}
\end{figure}

\begin{figure}[H]
  \centering
  \begin{minipage}[b]{0.4\textwidth}
    \includegraphics[width=1.1\textwidth]{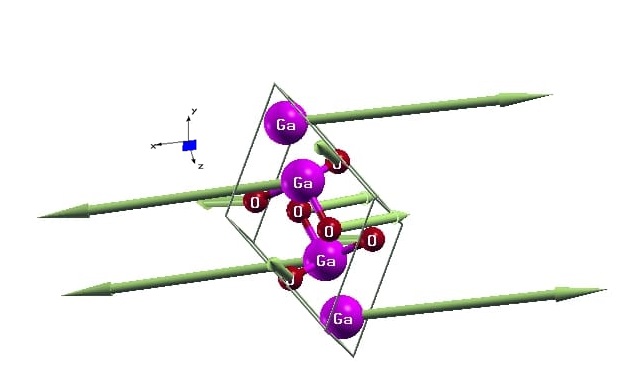}
    \caption{The net dipole moment of this mode is in the \textbf{x} cartesian direction $E_{u}^{x}$}
    \label{fig:Eux}
  \end{minipage}
  \hfill
  \begin{minipage}[b]{0.4\textwidth}
    \includegraphics[width=.5\textwidth]{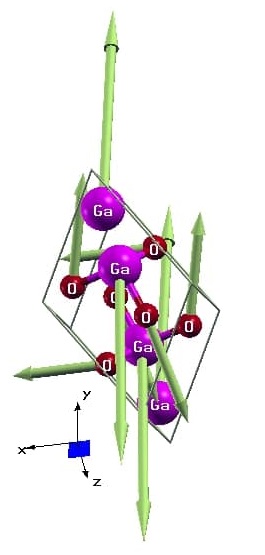}
    \caption{The net dipole moment of this mode is in the \textbf{y} cartesian direction $E_{u}^{y}$}
    \label{fig:Euy}
  \end{minipage}
\end{figure}

\begin{figure}[H]
    \centering
    \includegraphics[width=.2\linewidth]{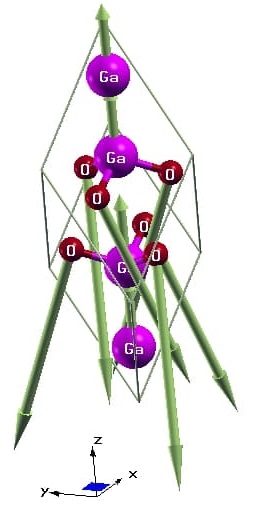}
    \caption{The net dipole moment of this mode is in the \textbf{z} cartesian direction $A_{2u}$}
    \label{fig:A2u}
\end{figure}

There are 27 optical modes in $\alpha-Ga_{2}O_{3}$ and the symmetry reduction results in 10 IR active modes as shown in the $fig\ref{fig:IR}$. Of the 10 IR modes, 4 modes are doubly degenerate with the mechanical representation as $2A_{2u} + 4E_{u}$. The net dipole moment of the 4(doubly degenerate) modes are polarized in the x and y cartesian directions ($E_{u}$) as shown in $fig\ref{fig:Eux}$ $\&$ $fig\ref{fig:Euy}$ respectively and the remaining 2 modes are polarized in the z direction ($A_{2u}$) shown in $fig\ref{fig:A2u}$. The directional polarization is captured in the DC dielectric constant and the LO-TO split. The strength of each of the IR active modes is directly proportional to the dipole moment and is calculated as the product of the Born effective charge and the atomic displacement pattern at the Brillouin zone centre. $Table\ref{tab:IRval}$ shows the comparison between the calculated IR active frequencies and the experimentally observed frequencies \cite{ref23}, showing a close match.

\FloatBarrier
\begin{table}[h]
\centering
\begin{tabular}{|c|c|}
\hline
\multicolumn{2}{|c|}{$\textbf{IR Modes }$ ($cm^{-1}$)} \\ \hline
$Calculated$ & $Experimental^{[25]}$ \\ \hline
\multicolumn{2}{|c|}{$E_{u}$} \\ \hline
220.88 & Not Observed \\ \hline
328.82 & 333.4 \\ \hline
462.21 & 469.9 \\ \hline
553.64 & 562.7 \\ \hline
\multicolumn{2}{|c|}{$A_{2u}$} \\ \hline
270.66 & 280 \\ \hline
535.11 & 544 \\ \hline
\end{tabular}
\captionof{table}{Comparison of the calculated IR frequencies with the experimental observations where the degenerate modes $E_{u}$ are polarized in the x and y cartesian directions and the $A_{2u}$ modes are polarized in the z cartesian direction}
\label{tab:IRval}
\end{table}
\FloatBarrier

\begin{figure}
\centering
\includegraphics[width=0.8\linewidth]{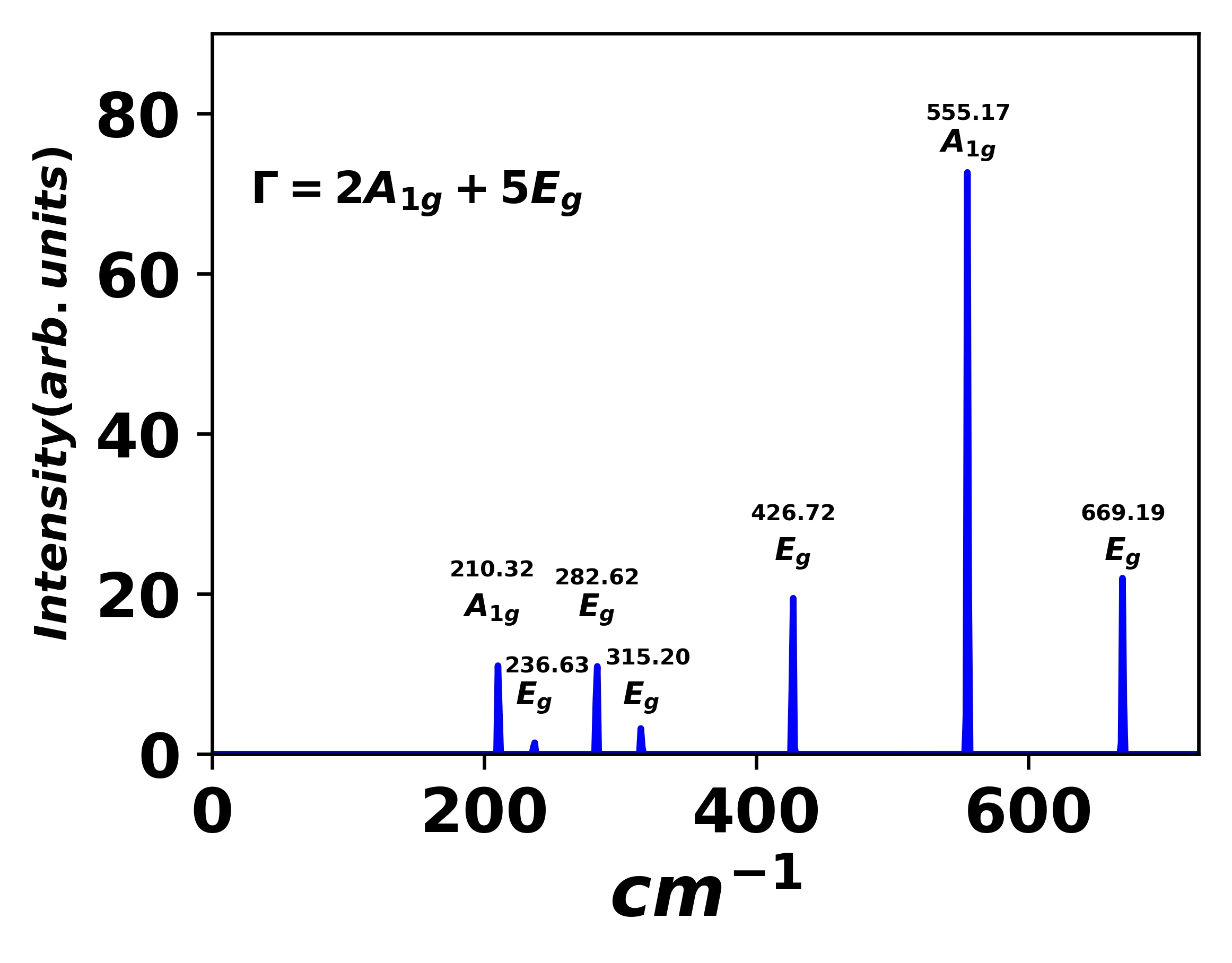}
\caption{The calculated Raman spectrum using the Lazzeri et.al formalism implemented in the QE package}
\label{fig:Raman}
\includegraphics[width=\linewidth]{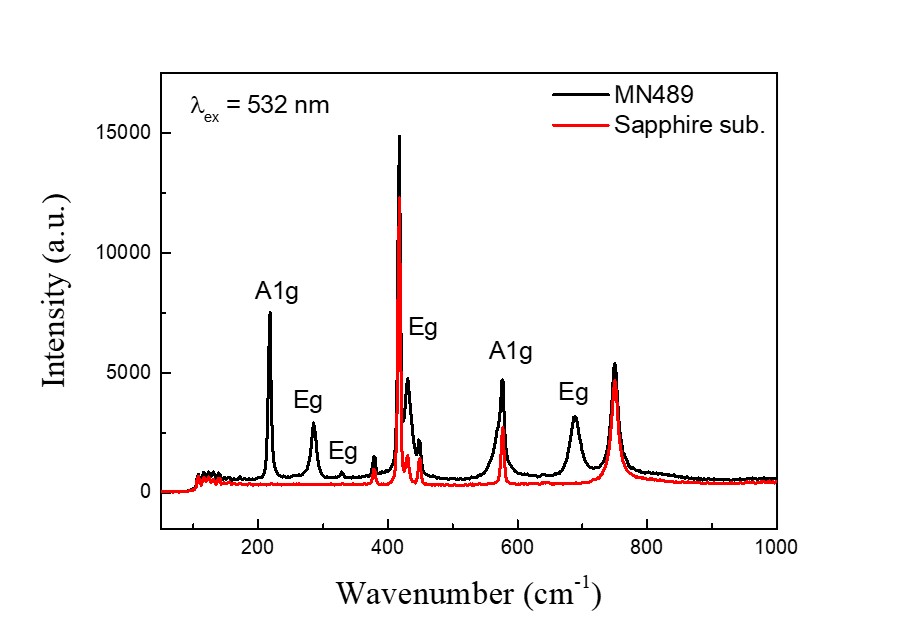}
\caption{The experimentally observed Raman spectrum obtained by our collaborators at U. Michigan}
\label{fig:Raman1}
\end{figure}

There are 12 Raman active modes in $\alpha-Ga_{2}O_{3}$ where 5 modes are doubly degenerate represented as $2A_{1g} + 5E_{g}$. The non-resonant Raman coefficients are calculated under the formalism proposed by Lazzeri \textit{et.al} \cite{ref24} and is implemented in the Quantum Espresso package. The calculated Raman spectra closely matches the spectra contained experimentally\cite{ref25} as shown in the $fig\ref{fig:Raman}$. The slight shift in the theoretical values compared to the experimental results may arise due to the dependence of the structural parameters on the pseudopotential used in the self-consistent calculation. $Table \ref{table:Ramanval}$ compares the theoretical Raman frequencies with the experimental observations.

\FloatBarrier
\begin{table}[h]
\centering
\begin{tabular}{|c|c|c|}
\hline
\multicolumn{3}{|c|}{$\textbf{Raman Modes}$ ($cm^{-1}$)} \\ \hline
$Calculated$ & $Experiment^{[27]}$ & $Experiment$ \\ \hline
\multicolumn{3}{|c|}{$E_{g}$} \\ \hline
236.63 & 240.7 & Not Observed \\ \hline
282.62 & 285.3 & 286 \\ \hline
315.2 & 328.8 & 328 \\ \hline
426.72 & 430.7 & 431 \\ \hline
669.19 & 686.7 & 688 \\ \hline
\multicolumn{3}{|c|}{$A_{1g}$} \\ \hline
210.32 & 218.2 & 218 \\ \hline
555.17 & 569.7 & 576 \\ \hline
\end{tabular}
\caption{Comparison of the calculated non-resonant Raman frequencies with the experimental observations. Experiment 2 corresponds to the spectrum shown in $fig\ref{fig:Raman1}$ where the first frequency is not observed}
\label{table:Ramanval}
\end{table}
\FloatBarrier

\newpage
\section{Scattering Rates}
The scattering rate is proportional to square of the EPI ($electron-phonon$) interaction elements which are calculated from the perturbation in the potential due to the lattice motion. The elements are the strength of transition from one state to another and are calculated by taking the overlap of the wavefunctions of the initial and final state in the presence of a perturbing potential. The generalized EPI element is given as:
\begin{equation}\label{epi}
g^{el-ph} = <\psi_{\textbf{k}+\textbf{q}}\mid \frac{\partial{V_{scf}}}{\partial{\textbf{q}}} \mid \psi_{\textbf{k}}>
\end{equation}

where $\psi$ is electronic wavefunction and $\partial{V_{scf}}$ is the perturbed potential due to lattice motion. We calculate the scattering rate using the Fermi's Golden rule. The q space integration in the Fermi's golden rule is carried out using the technique of Gaussian smearing for the energy conservation $\delta$ function. In the present calculation, we have used the parabolic band approximation with an isotropic effective mass. The $fig\ref{fig:TotScat}$ shows the polar optical phonon scattering, the non-polar scattering and the ionized impurity scattering rate calculated at 300K. We have taken the donor density of $1e^{17} cm^{-3}$ with donor activation energy of $31 meV$ \cite{ref11} assuming Silicon as the donor material.
\subsection{Polar Optical Phonon Scattering}
\begin{figure}[H]
\centering
\includegraphics[width=0.8\linewidth]{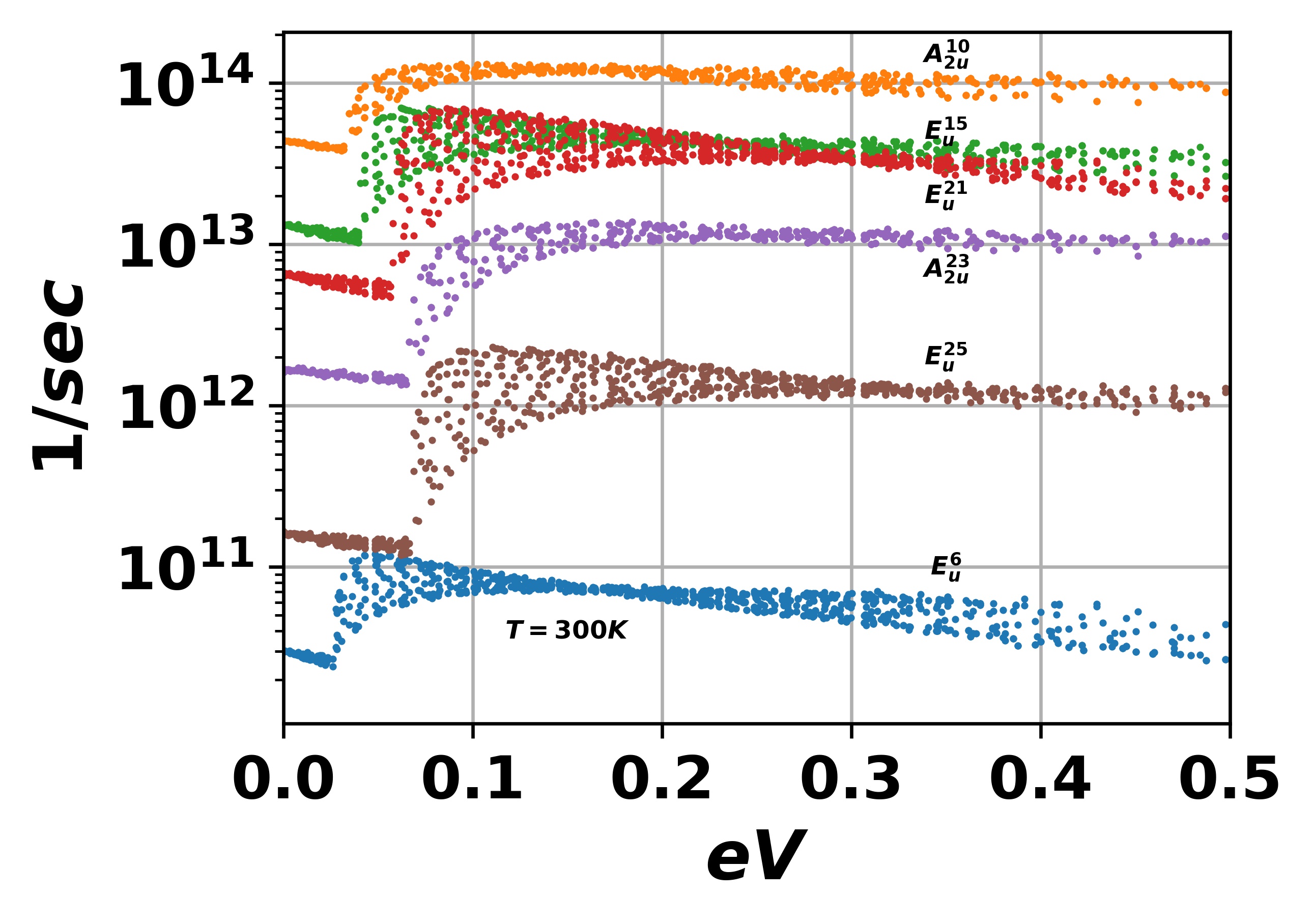}
\caption{Mode-wise Polar optical phonon scattering rate considering no phonon dispersion and parabolic band approximation with the 33 meV mode contributing maximum to the scattering rate. Also, sharp changes in the slope of the rates corresponding to different emission energies for each mode can be observed}
\label{fig:Modewise}
\end{figure}

\begin{figure}[H]
\centering
\includegraphics[width=0.8\linewidth]{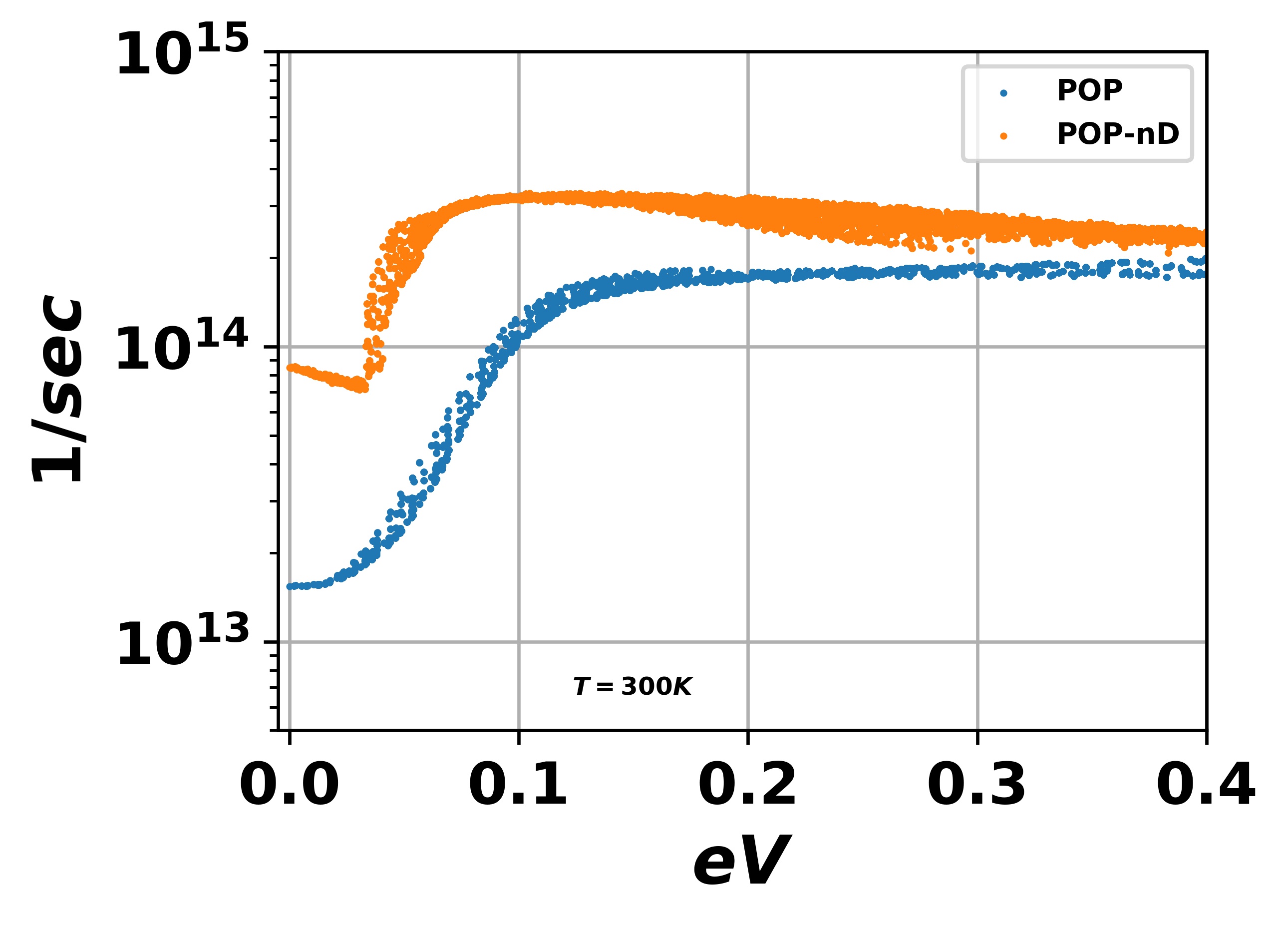}
\caption{The comparison of the POP scattering rates considering full phonon dispersion (blue) and no dispersion (orange). Clearly, the no dispersion scattering rate is high because moving away from the $\Gamma$ point reduces the net dipole moment}
\label{fig:DvsnD}
\end{figure}

To calculate the polar optical phonon EPI elements, we use the Vogl model \cite{ref28} under two different schemes. The Frohlich model that is widely used to estimate the EPI elements has a major drawback; it does not consider the anisotropy in the dipole moment and assumes the LO modes to be dispersion less. The Frohlich interaction elements are given by:
\begin{equation}
g_{\textbf{q}\nu}^{el-ph} = \frac{i}{\mid \textbf{q} \mid}\sqrt{\frac{e^2\hbar\omega_{\nu}}{2\epsilon_{0}\Omega}(\frac{1}{\epsilon_{\infty}}-\frac{1}{\epsilon_{s}})}
\end{equation} 

where $\omega$ is the phonon frequency of a given mode at the $\Gamma$ point, $\Omega$ is the normalization volume, $\epsilon_{\infty}$ is the spatially averaged high frequency dielectric constant and $\epsilon_{s}$ is the static dielectric constant. From the equation, it is clear that taking the spatial average of the dielectric constant leads to the averaging of the anisotropic effects and also it does not take into consideration the polarization direction of each mode of the phonon spectrum.
\newline
\newline
To incorporate the anisotropy in the EPI elements, we use the Vogl model. In the first scheme, we still assume the phonon spectrum to be dispersion less, but we take the projection of the dipole moment for each mode onto the phonon wavevector thus accounting for the anisotropic LO-TO split. The expression for the EPI elements used under this scheme is given by \cite{ref29}:
\begin{equation}
g_{\textbf{q}\nu} = i\frac{e^2}{\Omega\epsilon_{0}}\sum_{j}\sqrt{\frac{\hbar}{2M_{j}\omega_{j\nu}}}\frac{\textbf{q}.Z_{eff}.e_{j\nu}}{\epsilon_{\infty}(\mid \textbf{q} \mid^{2} + q_{scr}^{2})}
\end{equation}

where, $Z_{eff}$ is the Born effective charge, $e_{\nu}$ is the phonon eigenvector for each atom in the unit cell at the $\Gamma$ point, $\omega_{\nu}$ is the phonon frequency of each mode, $M_{j}$ is the atomic mass of each of the constituent atoms and $q_{scr}$ is the Thomas-Fermi screening wavevector. The inclusion of the screening term circumvents the divergence at $q=0$ \cite{ref29}. The high frequency dielectric constant is nearly spatially isotropic; hence we take the average of the trace of the dielectric tensor. The screening wavevector is calculated from the Fermi energy and is expressed as:
\begin{equation}
q_{scr} = \sqrt{\frac{6\pi n e^{2}}{\epsilon E_{f}}}
\end{equation}

where, $E_{f}$ is the Fermi energy and $n$ is the charge density. One of the drawbacks of this assumption is the approximation of the dispersion less phonon spectrum. This approximation overestimates the scattering rate when compared to the full phonon dispersion, as can be seen in the $fig\ref{fig:DvsnD}$. The net strength of the dipole moment decreases moving away from the $\Gamma$ point and hence the contribution from the $q$ point to the scattering rate also decreases. However, this is not the case when we assume dispersion less phonon spectrum. The EPI elements in this case only decay because of the increase in the magnitude of the phonon wavevector as the dipole moment remains constant. From the plot of the IR spectrum in $fig\ref{fig:IR}$, we see that the $10^{th}$ (270.66 $cm^{-1}$) mode has the maximum IR strength and hence is expected to contribute most to the scattering rate. Moreover, the phonon occupancy of this mode (270.66 $cm^{-1}$) $33 meV$ is high at room temperature which will again play a major role in determining the dominant scattering mode. The mode dependent plot of the bulk scattering rate in $fig\ref{fig:Modewise}$ clearly shows the dominant mode. Also, from the figure, we can see sharp emissions depicted by the sudden change in slope, depending on the energy of each mode. Another noteworthy point is that the degenerate energy levels have different scattering rates arising due to the anisotropy which is averaged out when calculating the electron mobility.
\newline
\newline
In the second scheme, we consider the full phonon dispersion, but we do not explicitly include any screening. We only use the high frequency dielectric constant tensor to screen the LO phonons arising due to the valence electrons. The generalized expression for the Vogl model for the polar optical phonon EPI elements proposed by Carla Verdi \textit{et.al}\cite{ref30} is given as:
\begin{equation}
\resizebox{.9\hsize}{!}{$g_{\nu}^{el-ph}(\textbf{k},\textbf{q}) = i\frac{4\pi}{\Omega}\frac{e^2}{4\pi\epsilon_{0}}\sum_{\kappa}(\frac{\hbar}{2NM_{\kappa}\omega_{q\nu}})^{\frac{1}{2}}\sum_{G\ne -q}\frac{(\textbf{q}+\textbf{G}).Z_{\kappa}.e_{\kappa\nu}(\textbf{q})}{(\textbf{q}+\textbf{G}).\epsilon_{\infty}.(\textbf{q}+\textbf{G})}e^{-i\tau_{\kappa}.(\textbf{q}+\textbf{G})}[U_{k+q}^{\dagger}U_{k}]$}
\end{equation}
where, $\tau$ is the atomic position, $\omega$ is the phonon frequency and $U$ are the unitary rotation matrices. The sum is over all the reciprocal lattice vectors $\textbf{G}$ and the number of atoms in the unit cell. WE have used 64000 k points and 68921 q points for the EPI calculation. The product of $U$ matrices accounts for the overlap of the wavefunction and can be obtained during the Wannier interpolation.  The polar optical phonon scattering rate is shown in the $fig\ref{fig:TotScat}$.

\subsection{Non Polar Scattering}
The non polar EPI elements are calculated using the EPW code \cite{ref31}. The total electron phonon interaction elements calculated using $equation\ref{epi}$, is split into the longrange and the shortrange component $g = g^{L} + g^{S}$ for interpolation from coarse grid onto a fine grid using the Wannier functions. The longrange part of the EPI elements cannot be interpolated using the Wannier functions because of the localized nature of the Wannier functions in the real space. 
\newline
\newline
The shortrange elements consist of deformation potential scattering elements (both optical and acoustic). Once the shortrange EPI elements are calculated on a coarse grid, the interpolation on to a fine grid is done in two steps: firstly, the EPI elements in the Wannier space is interpolated onto a fine reciprocal space phonon wavevector grid \textit{i.e} phonon Bloch representation. The resulting interaction elements are now in the electronic Wannier space and fine phonon Bloch space. We extract the EPI's at this stage and then use the Fourier interpolation to calculate the elements on a fine electronic Bloch space given as \cite{ref32}:
\begin{equation}
g(\textbf{k},\textbf{q}) = \frac{1}{N_{e}}\sum_{R_{e}}e^{i\textbf{k}.\textbf{Re}}U_{\textbf{k}+\textbf{q}}g(\textbf{Re},\textbf{q})U_{\textbf{k}}^{\dagger}
\end{equation}

where $g(\textbf{k},\textbf{q})$ are the EPI elements on a fine electron and phonon reciprocal space grid,  $N_{e}$ is the degeneracy of the Wigner-Seitz cell, $R_{e}$ is the Wigner-Seitz cell center, $g(R_{e},q)$ is the EPI elements on a fine phonon grid in Bloch space and electronic Wannier space. The unitary rotation matrices are unity in our case, since the number of bands inside the energy window used during the process of Wannierization \cite{ref20, ref21} is just one \textit{i.e} the lowermost conduction band. Such band mixing is done to remove non-analyticity in the bands which may arise due to the band crossing making the maximal localization of the Wannier functions challenging.
\newline
\newline
To calculate the scattering rate, we use the Fermi's Golden rule, by interpolating the EPI elements on a fine k and q point grid on the fly and using Gaussian smearing for energy conservation. For the shortrange scattering rate calculation, we have used around 45000 q points and 55000 k points distributed uniformly inside the Brillouin zone. We have covered only $15\%$ of the Brillouin zone, because large phonon wavevectors will not contribute to the scattering process due to the unavailability of equivalent valleys to account for such a large momentum change. 

\subsection{Ionized Impurity Scattering}
\begin{figure}[H]
\centering
\includegraphics[width=0.8\linewidth]{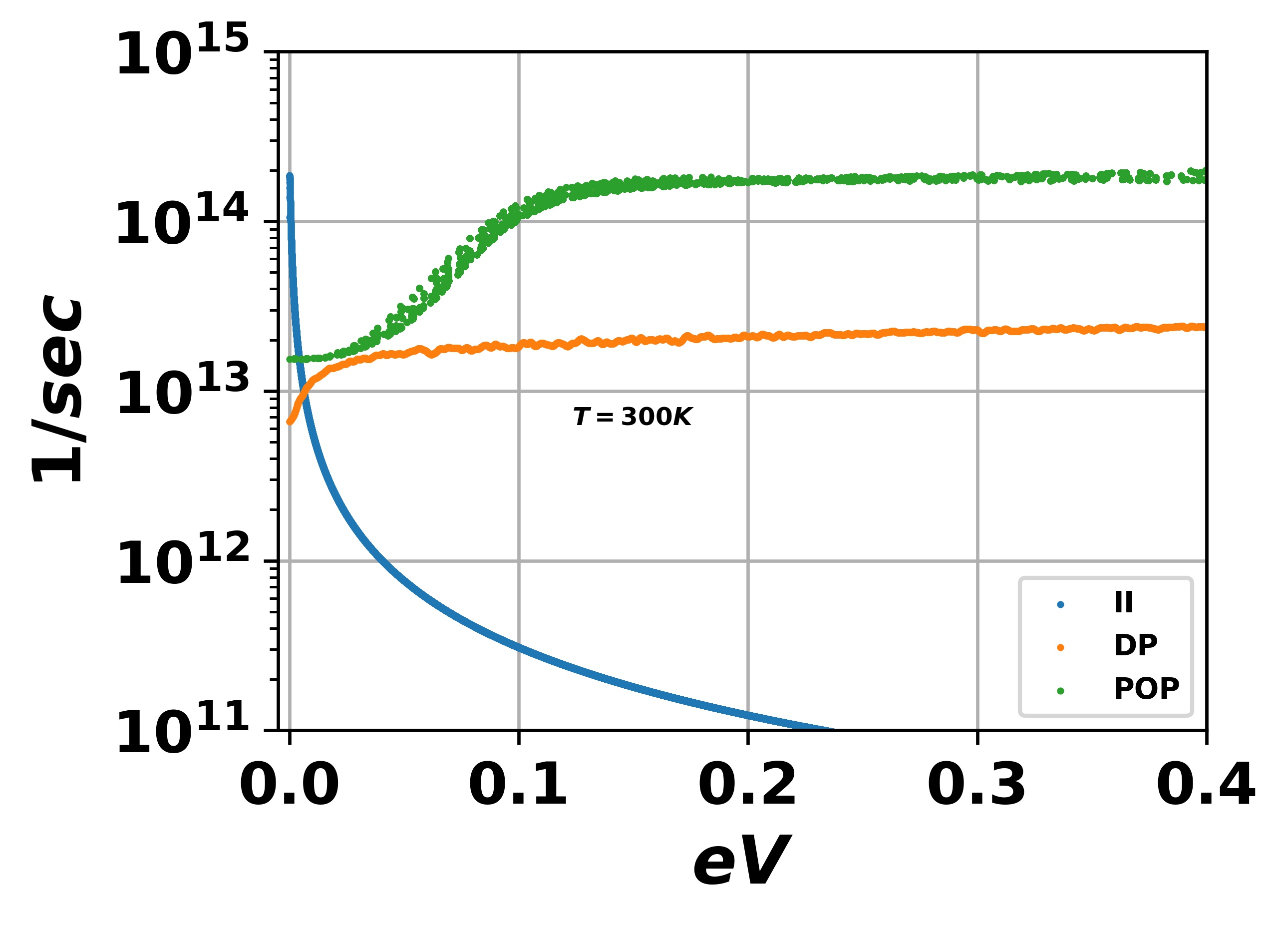}
\caption{The green, orange and the blue curves show the 3 different scattering mechanisms- POP, DP and II respectively at 300K. The POP scattering is the dominant mechanism}
\label{fig:TotScat}
\end{figure}

The ionized impurity scattering rate is calculated using the Brooks Herring model \cite{ref26} with a constant Debye like screening. The analytic expression for the momentum relaxation time for the ionized impurity scattering is given by \cite{ref27}:
\begin{equation}
\tau_{m}(p) = \frac{16*\sqrt{2m^{*}}\pi\kappa_{s}^{2}\epsilon_{0}^{2}}{N_{I}q^{4}}[\log{(1+\gamma^{2})-\frac{\gamma^{2}}{1+\gamma^{2}}}]^{2}E^{\frac{3}{2}}(p)
\end{equation} 

where $\gamma$ is the function of the Debye length, $\kappa$ is the DC dielectric constant, $p$ is the electron momentum and $N_{I}$ is the total impurity concentration. In $fig\ref{fig:TotScat}$ we can see that at very low energies, the ionized impurity scattering is the dominant mechanism and it falls off rapidly with the increase in the electron energy since the duration of interaction of the screened field with the electron decreases. We have used the spatially averaged DC dielectric constant, thus loosing information about anisotropy that arises due to the LO-TO split as is evident from $table\ref{table:dielectric}$.

\section{Electron Mobility}
In this study, we use the relaxation time approximation to calculate the electron mobility which is given by:
\begin{equation}
\mu = \frac{e<<\tau>>}{m}
\end{equation}

where, $\mu$ is the electron mobility in $cm^{2}/Vs$, $e$ is the electronic charge, $m$ is the electron effective mass and $\tau$ is the average relaxation time. The average value is calculated as \cite{ref33}:
\begin{equation}
<<\tau>> = \frac{\int_{0}^{E}\frac{\partial{f_{0}}}{\partial{E}}\tau(E)E^{\frac{3}{2}}}{\int_{0}^{E}\frac{\partial{f_{0}}}{\partial{E}}E^{\frac{3}{2}}}
\end{equation}

where, $f_{0}$ is the equilibrium Fermi-Dirac distribution, $E$ is the electron energy and $\tau$ is the relaxation time. We calculate the mobility for all the three different scattering mechanism we have accounted for and then use the Matthiessen's rule to calculate the overall electron mobility which is given as:
\begin{equation}
\frac{1}{\mu} = \frac{1}{\mu_{POP}} + \frac{1}{\mu_{NP}} + \frac{1}{\mu_{II}}
\end{equation}

where, $\mu_{POP}$, $\mu_{NP}$ and $\mu_{II}$ is the mobility due to the polar optical phonon, non-polar and ionized impurity scattering respectively. We do this calculation at $300K$ and the bulk electron mobility obtained is about $174 cm^{2}/Vs$. This is dominated by the POP ($300 cm^{2}/Vs$)and non-polar phonon ($420 cm^{2}/Vs$) as compared to the ionized impurity scattering. These calcualted values are just the first order approximation to the electron mobility. Using better approximation to solve the Boltzmann transport equation would be through the Rode's iterative technique can increase the mobility by around $30-40\%$ . Also the inclusion of screening will result in the further increase of electron mobility.

\section{Conclusion}
We have theoretically investigated the electron-phonon coupling in $\alpha-Ga_{2}O_{3}$ and estimated the low field electron mobility at 300K. We take into account the polar optical phonon, ionized impurity and deformation potential scattering for the mobility calculation. The polar optical phonon interaction elements were estimated assuming dispersion less phonon spectrum and also with the full phonon dispersion. The DC dielectric constant was calculated using LST relation which clearly shows strong anisotropy in the z cartesian direction. The infrared and the Raman spectrum was calculated and compared with the experiments showing close match.

\newpage
\section{Acknowledgement}
We acknowledge the support from Air Force Office of Scientific Research
under award number FA9550-18-1-0479 (Program Manager: Ali Sayir) and
from NSF under awards  ECCS 1607833 and ECCS - 1809077. We also acknowledge the high-performance computing cluster provided by The Centre for Computational Research at the University at Buffalo.

\newpage
\bibliography{References}

\begin{thebibliography}{10}

\bibitem{ref1}
Rustum Roy, V.~G. Hill, and E.~F. Osborn.
\newblock Polymorphism of ${Ga}_{2}{O}_{3}$ and the system
  ${Ga}_{2}{O}_{3}-{H}_{2}{O}$.
\newblock {\em Journal of the American Chemical Society}, 74(3):719--722, 1952.

\bibitem{ref2}
H.~Y. Playford, A.~C. Hannon, E.~R. Barney, and R.~I. Walton.
\newblock {{S}tructures of uncharacterised polymorphs of gallium oxide from
  total neutron diffraction}.
\newblock {\em Chemistry}, 19(8):2803--2813, Feb 2013.

\bibitem{ref3}
Haiying He, Roberto Orlando, Miguel~A. Blanco, Ravindra Pandey, Emilie
  Amzallag, Isabelle Baraille, and Michel R\'erat.
\newblock First-principles study of the structural, electronic, and optical
  properties of ${Ga}_{2}{O}_{3}$ in its monoclinic and hexagonal phases.
\newblock {\em Phys. Rev. B}, 74:195123, Nov 2006.

\bibitem{ref4}
Masataka Higashiwaki, Kohei Sasaki, Akito Kuramata, Takekazu Masui, and
  Shigenobu Yamakoshi.
\newblock Development of gallium oxide power devices.
\newblock {\em physica status solidi (a)}, 211(1):21--26, 2014.

\bibitem{ref5}
Masataka Higashiwaki, Kohei Sasaki, Hisashi Murakami, Yoshinao Kumagai, Akinori
  Koukitu, Akito Kuramata, Takekazu Masui, and Shigenobu Yamakoshi.
\newblock Recent progress in ${Ga}_{2}{O}_{3}$ power devices.
\newblock {\em Semiconductor Science and Technology}, 31(3):034001, jan 2016.

\bibitem{ref6}
K.~{Zeng}, A.~{Vaidya}, and U.~{Singisetti}.
\newblock 1.85 kv breakdown voltage in lateral field-plated ${Ga}_{2}{O}_{3}$
  mosfets.
\newblock {\em IEEE Electron Device Letters}, 39(9):1385--1388, Sep. 2018.

\bibitem{ref7}
K.~{Sasaki}, M.~{Higashiwaki}, A.~{Kuramata}, T.~{Masui}, and S.~{Yamakoshi}.
\newblock $\hbox{Ga}_{2} \hbox{O}_{3}$ schottky barrier diodes fabricated by
  using single-crystal $\beta$– $\hbox{Ga}_{2} \hbox{O}_{3}$ (010)
  substrates.
\newblock {\em IEEE Electron Device Letters}, 34(4):493--495, April 2013.

\bibitem{ref8}
Toshiyuki Oishi, Yuta Koga, Kazuya Harada, and Makoto Kasu.
\newblock High-mobility ${Ga}_{2}{O}_{3}$ single crystals grown by edge-defined
  film-fed growth method and their schottky barrier diodes with ${Ni}$ contact.
\newblock {\em Applied Physics Express}, 8(3):031101, feb 2015.

\bibitem{ref9}
M.~H. {Wong}, K.~{Goto}, H.~{Murakami}, Y.~{Kumagai}, and M.~{Higashiwaki}.
\newblock Current aperture vertical $\beta-{Ga}_{2}{O}_{3}$ mosfets fabricated
  by ${N}$- and ${Si}$-ion implantation doping.
\newblock {\em IEEE Electron Device Letters}, 40(3):431--434, March 2019.

\bibitem{ref10}
Yi~Lu, H.~H. Yao, Jingtao Li, Jianchang Yan, Junxi Wang, Jinmin Li, and
  Xiaohang Li.
\newblock Aln/beta-ga2o3 based hemt: a potential pathway to ultimate high power
  device.
\newblock 2019.

\bibitem{ref11}
Krishnendu Ghosh and Uttam Singisetti.
\newblock Ab initio calculation of electron–phonon coupling in monoclinic
  $\beta-{Ga}_{2}{O}_{3}$ crystal.
\newblock {\em Applied Physics Letters}, 109(7):072102, 2016.

\bibitem{ref12}
J.W. Roberts, P.R. Chalker, B.~Ding, R.A. Oliver, J.T. Gibbon, L.A.H. Jones,
  V.R. Dhanak, L.J. Phillips, J.D. Major, and F.C.-P. Massabuau.
\newblock Low temperature growth and optical properties of
  $\alpha-{Ga}_{2}{O}_{3}$ deposited on sapphire by plasma enhanced atomic
  layer deposition.
\newblock {\em Journal of Crystal Growth}, 528:125254, 2019.

\bibitem{ref35}
J.~H. Leach, K.~Udwary, J.~Rumsey, G.~Dodson, H.~Splawn, and K.~R. Evans.
\newblock Halide vapor phase epitaxial growth of $\beta-{Ga}_{2}{O}_{3}$ and
  $\alpha-{Ga}_{2}{O}_{3}$ films.
\newblock {\em APL Materials}, 7(2):022504, 2019.

\bibitem{ref13}
Daisuke Shinohara and Shizuo Fujita.
\newblock Heteroepitaxy of corundum-structured $\alpha-{Ga}_{2}{O}_{3}$ thin
  films on $\alpha-{Al}_{2}{O}_{3}$ substrates by ultrasonic mist chemical
  vapor deposition.
\newblock {\em Japanese Journal of Applied Physics}, 47(9):7311--7313, Sep
  2008.

\bibitem{ref14}
Krishnendu Ghosh and Uttam Singisetti.
\newblock Theory of high field transport in $\beta-{Ga}_{2}{O}_{3}$.
\newblock {\em International Journal of High Speed Electronics and Systems},
  28(01n02):1940008, 2019.

\bibitem{ref17}
P~Giannozzi, O~Andreussi, T~Brumme, O~Bunau, M~Buongiorno Nardelli, M~Calandra,
  R~Car, C~Cavazzoni, D~Ceresoli, M~Cococcioni, N~Colonna, I~Carnimeo, A~Dal
  Corso, S~de~Gironcoli, P~Delugas, R~A~DiStasio Jr, A~Ferretti, A~Floris,
  G~Fratesi, G~Fugallo, R~Gebauer, U~Gerstmann, F~Giustino, T~Gorni, J~Jia,
  M~Kawamura, H-Y Ko, A~Kokalj, E~Küçükbenli, M~Lazzeri, M~Marsili,
  N~Marzari, F~Mauri, N~L Nguyen, H-V Nguyen, A~Otero de-la Roza, L~Paulatto,
  S~Poncé, D~Rocca, R~Sabatini, B~Santra, M~Schlipf, A~P Seitsonen,
  A~Smogunov, I~Timrov, T~Thonhauser, P~Umari, N~Vast, X~Wu, and S~Baroni.
\newblock Advanced capabilities for materials modelling with quantum espresso.
\newblock {\em Journal of Physics: Condensed Matter}, 29(46):465901, 2017.

\bibitem{ref18}
Paolo Giannozzi, Stefano Baroni, Nicola Bonini, Matteo Calandra, Roberto Car,
  Carlo Cavazzoni, Davide Ceresoli, Guido~L Chiarotti, Matteo Cococcioni,
  Ismaila Dabo, Andrea {Dal Corso}, Stefano de~Gironcoli, Stefano Fabris, Guido
  Fratesi, Ralph Gebauer, Uwe Gerstmann, Christos Gougoussis, Anton Kokalj,
  Michele Lazzeri, Layla Martin-Samos, Nicola Marzari, Francesco Mauri,
  Riccardo Mazzarello, Stefano Paolini, Alfredo Pasquarello, Lorenzo Paulatto,
  Carlo Sbraccia, Sandro Scandolo, Gabriele Sclauzero, Ari~P Seitsonen,
  Alexander Smogunov, Paolo Umari, and Renata~M Wentzcovitch.
\newblock Quantum espresso: a modular and open-source software project for
  quantum simulations of materials.
\newblock {\em Journal of Physics: Condensed Matter}, 21(39):395502 (19pp),
  2009.

\bibitem{ref15}
Anton Kokalj.
\newblock Xcrysden—a new program for displaying crystalline structures and
  electron densities.
\newblock {\em Journal of Molecular Graphics and Modelling}, 17(3):176 -- 179,
  1999.

\bibitem{ref16}
Koichi Momma and Fujio Izumi.
\newblock {{\it VESTA3} for three-dimensional visualization of crystal,
  volumetric and morphology data}.
\newblock {\em Journal of Applied Crystallography}, 44(6):1272--1276, Dec 2011.

\bibitem{ref34}
M.~Marezio and J.~P. Remeika.
\newblock Bond lengths in the $\alpha-{Ga}_{2}{O}_{3}$ structure and the
  high‐pressure phase of ${Ga}_{2}-{x}{Fe}_{x}{O}_{3}$.
\newblock {\em The Journal of Chemical Physics}, 46(5):1862--1865, 1967.

\bibitem{ref19}
Gregory~H. Wannier.
\newblock The structure of electronic excitation levels in insulating crystals.
\newblock {\em Phys. Rev.}, 52:191--197, Aug 1937.

\bibitem{ref20}
Nicola Marzari and David Vanderbilt.
\newblock Maximally localized generalized wannier functions for composite
  energy bands.
\newblock {\em Phys. Rev. B}, 56:12847--12865, Nov 1997.

\bibitem{ref21}
Nicola Marzari, Arash~A. Mostofi, Jonathan~R. Yates, Ivo Souza, and David
  Vanderbilt.
\newblock Maximally localized wannier functions: Theory and applications.
\newblock {\em Rev. Mod. Phys.}, 84:1419--1475, Oct 2012.

\bibitem{ref22}
Xavier Gonze and Changyol Lee.
\newblock Dynamical matrices, born effective charges, dielectric permittivity
  tensors, and interatomic force constants from density-functional perturbation
  theory.
\newblock {\em Phys. Rev. B}, 55:10355--10368, Apr 1997.

\bibitem{ref23}
Martin Feneberg, Jakob Nixdorf, Maciej~D. Neumann, Norbert Esser, Lluis
  Art\'us, Ramon Cusc\'o, Tomohiro Yamaguchi, and R\"udiger Goldhahn.
\newblock Ordinary dielectric function of corundumlike $\beta-{Ga}_{2}{O}_{3}$
  from 40 mev to 20 ev.
\newblock {\em Phys. Rev. Materials}, 2:044601, Apr 2018.

\bibitem{ref24}
Michele Lazzeri and Francesco Mauri.
\newblock First-principles calculation of vibrational raman spectra in large
  systems: Signature of small rings in crystalline $sio_{2}$.
\newblock {\em Phys. Rev. Lett.}, 90:036401, Jan 2003.

\bibitem{ref25}
R.~Cuscó, N.~Domènech-Amador, T.~Hatakeyama, T.~Yamaguchi, T.~Honda, and
  L.~Artús.
\newblock Lattice dynamics of a mist-chemical vapor deposition-grown
  corundum-like ${Ga}_{2}{O}_{3}$ single crystal.
\newblock {\em Journal of Applied Physics}, 117(18):185706, 2015.

\bibitem{ref28}
P.~Vogl.
\newblock Microscopic theory of electron-phonon interaction in insulators or
  semiconductors.
\newblock {\em Phys. Rev. B}, 13:694--704, Jan 1976.

\bibitem{ref29}
Youngho Kang, Karthik Krishnaswamy, Hartwin Peelaers, and Chris G~Van de~Walle.
\newblock Fundamental limits on the electron mobility of
  $\beta-{Ga}_{2}{O}_{3}$.
\newblock {\em Journal of Physics: Condensed Matter}, 29(23):234001, May 2017.

\bibitem{ref30}
Carla Verdi and Feliciano Giustino.
\newblock Frohlich electron-phonon vertex from first principles.
\newblock {\em Phys. Rev. Lett.}, 115:176401, Oct 2015.

\bibitem{ref31}
S.~Poncé, E.R. Margine, C.~Verdi, and F.~Giustino.
\newblock Epw: Electron–phonon coupling, transport and superconducting
  properties using maximally localized wannier functions.
\newblock {\em Computer Physics Communications}, 209:116 -- 133, 2016.

\bibitem{ref32}
Feliciano Giustino, Marvin~L. Cohen, and Steven~G. Louie.
\newblock Electron-phonon interaction using wannier functions.
\newblock {\em Phys. Rev. B}, 76:165108, Oct 2007.

\bibitem{ref26}
D.~Chattopadhyay and H.~J. Queisser.
\newblock Electron scattering by ionized impurities in semiconductors.
\newblock {\em Rev. Mod. Phys.}, 53:745--768, Oct 1981.

\bibitem{ref27}
Mark Lundstrom.
\newblock {\em Fundamentals of Carrier Transport}.
\newblock Cambridge University Press, 2 edition, 2000.

\bibitem{ref33}
Antonella Parisini and Roberto Fornari.
\newblock Analysis of the scattering mechanisms controlling electron mobility
  in crystals.
\newblock {\em Semiconductor Science and Technology}, 31(3):035023, Feb 2016.

\end{thebibliography}
\bibliographystyle{unsrt}
\end{document}